\documentclass{camera}
\usepackage{graphicx}
\usepackage{multirow}

\begin{document}

\title{Rising indications for three\\
gamma-ray burst groups}

\author{J. \v{R}\'{\i}pa$^{1}$, D. Huja$^{1}$, A.~M\'{e}sz\'{a}ros$^{1}$ \and C. Wigger$^{2}$}

\organization{
$^{1}$ Charles University, Faculty of Mathematics and Physics,\\
Astronomical Institute, Prague, Czech Republic\\
$^{2}$ Kantonsschule Wohlen, Switzerland}

\maketitle

\begin{abstract}
Several papers were written about the gamma-ray burst (GRBs)
groups. Our statistical study is based on the durations and
hardness ratios of the \emph{Swift} and \emph{RHESSI} data.
\end{abstract}

The \emph{RHESSI} sample consists of 472 GRBs from Feb. 2002 to Feb.
2009. When the hardness ratios are used in our analysis, we
took only bursts before Nov. 2007, \emph{i.e.}, before the
spectrometer's annealing (425 GRBs). The \emph{Swift} sample consists
of 388 GRBs from Nov. 2004 to Dec. 2007.

There is a rising number of the statistical works giving
results that three groups may exist (BATSE \cite{ref01}, \emph{Swift}
\cite{ref02,ref03}, \emph{Beppo-SAX} \cite{ref04} and \emph{RHESSI}
\cite{ref05} data). We analyse the updated \emph{RHESSI} and \emph{Swift}
databases, use the minimal $\chi^2$ fitting of the $T_{90}$
histogram and F-test, the maximum likelihood (ML) method and ML
ratio test applied on the $T_{90}$ durations alone (1D) or
$T_{90}$/hardnesses $H$ pairs (2D) (see fig.~1). We assume that
$T_{90}$ and $H$ are log-normally distributed.

Summarizing our results, we confirm, from the statistical point
of view, a significant ($P<5\,\%$) improvement in the $\chi^2$ and
$L$ values if a third group is concerned (see table~1).
\newline

We acknowledge the support from the grants GAUK 46307, GA\v{C}R
205/08/H005, MSM0021620860, and OTKA K077795.
\nopagebreak
\begin{table} \caption{
The $\chi^2$ and ML methods. For the \emph{RHESSI} data, 3a and 3b are
both 3 log-normal fits, but with different properties. $P$ is
the probability (in \%) that the improvement, when 3 functions
are used, is accidental (given by F or ML ratio tests). $gof$
is the goodness of fit (in \%).} \centering
\begin{tabular}{cccccccccccc}

\cline{1-7} \cline{9-12}
\\[-2.8ex]
                                        &                                         & fit & $dof$ & $\chi^2$ & $gof$ & $P_{\rm F-test}$ & &                                        &                                       &  $L$   & $P_{\rm ML-test}$\\
\\[-2.8ex]
\cline{1-7} \cline{9-12}
\\[-2.5ex]
\cline{1-7} \cline{9-12}
\\[-2.7ex]
\multirow{3}{*}{\rotatebox{90}{\emph{Swift}}} &\multirow{3}{*}{\rotatebox{90}{$\chi^2$}} & 2   & 20    &   17.0   & 59.0  &              & &\multirow{3}{*}{\rotatebox{90}{\emph{RHESSI}}} &\multirow{3}{*}{\rotatebox{90}{ML 1D}} & -438.6 &              \\
                                       &                                          & 3a  & 17    &   10.0   & 86.7  & \textbf{2.5} & &                                        &                                       & -433.8 & \textbf{2.3} \\
                                       &                                          & 3b  &       &          &       &              & &                                        &                                       & -434.2 & \textbf{3.3} \\
\\[-2.8ex]
\cline{1-7} \cline{9-12}
\\[-2.8ex]
\multirow{3}{*}{\rotatebox{90}{\emph{RHESSI}}} &\multirow{3}{*}{\rotatebox{90}{$\chi^2$}} & 2  & 23    &   31.9   &  7.9  &              & &\multirow{3}{*}{\rotatebox{90}{\emph{RHESSI}}} &\multirow{3}{*}{\rotatebox{90}{ML 2D}} & -318.8 &              \\
                                        &                                          & 3a & 20    &   23.2   & 23.0  & 8.7          & &                                        &                                       & -310.0 & \textbf{0.7} \\
                                        &                                          & 3b & 20    &   20.0   & 39.8  & \textbf{2.2} & &                                        &                                       & -311.7 & \textbf{2.7} \\
\\[-2.8ex]
\cline{1-7} \cline{9-12}
\end{tabular}
\end{table}

\begin{figure}
\begin{center}
\scalebox{0.6}{\includegraphics*[0,0][595,267]{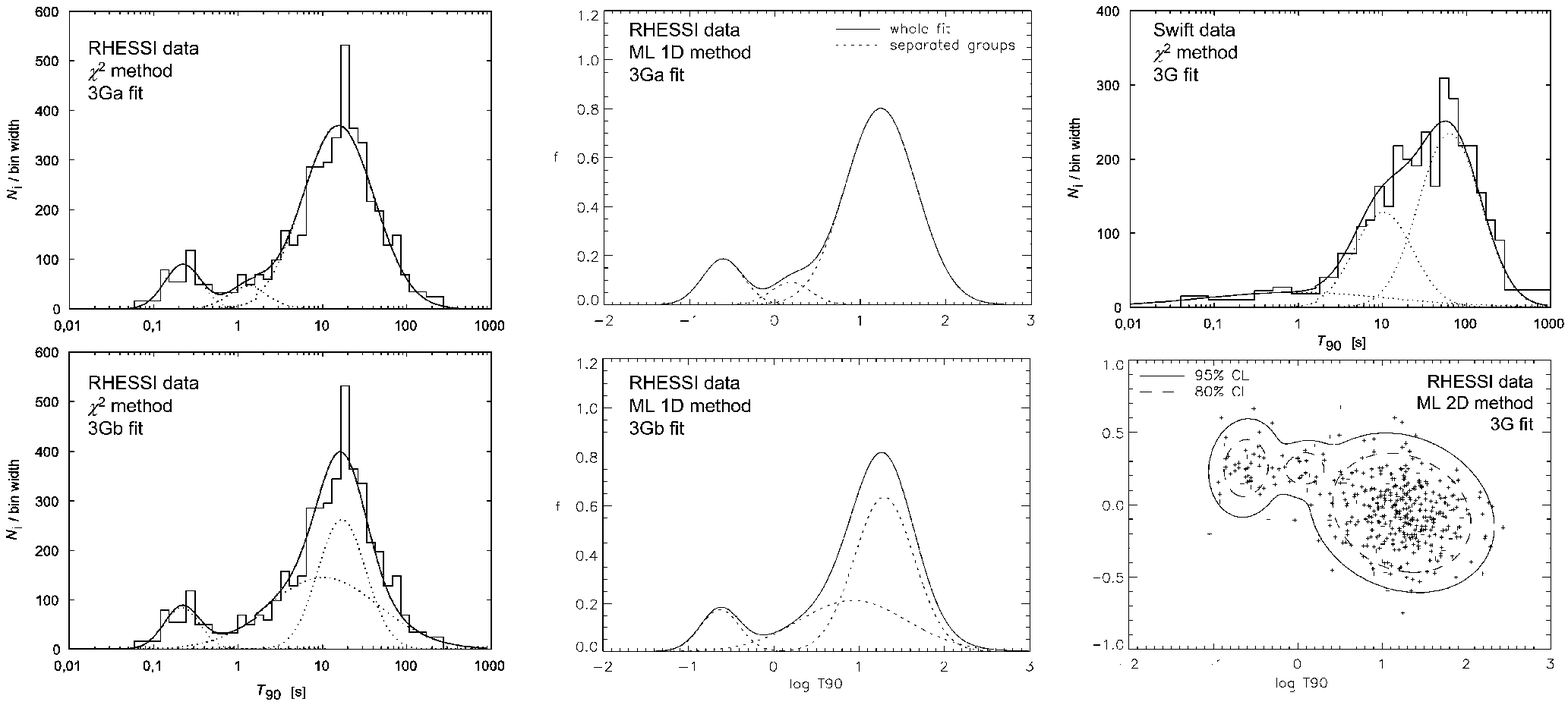}}
\caption{The best $\chi^2$ and ML fits on the \emph{RHESSI} and \emph{Swift} data.}
\end{center}
\end{figure}


\begin{thebibliography}{}
\bibitem[1]{ref01} HORV\'{A}TH I. ET AL., \emph{A\&A}, \textbf{447} (2006) 23.
\bibitem[2]{ref02} HORV\'{A}TH I. ET AL., \emph{A\&A}, \textbf{489} (2008) L1.
\bibitem[3]{ref03} HUJA D. ET AL., \emph{A\&A}, \textbf{504} (2009) 67.
\bibitem[4]{ref04} HORV\'{A}TH I., \emph{Ap\&SS}, \textbf{323} (2009) 83.
\bibitem[5]{ref05} {\v R}{\'I}PA J. ET AL., \emph{A\&A}, \textbf{498} (2009) 399.
\end{thebibliography}
\end{document}